\documentclass[]{aastex631}

\usepackage{comment,bm}
\shortauthors{French et al.}

\graphicspath{{./}{figures/}}

\accepted{to ApJL, February 3rd 2023}

\begin{document}

\title{First Observation of Chromospheric Waves in a Sunspot by DKIST/ViSP: The Anatomy of an Umbral Flash}


\author[0000-0001-9726-0738]{Ryan J. French}
\affiliation{National Solar Observatory, 3665 Innovation Drive, Boulder CO 80303}
\affiliation{Mullard Space Science Laboratory,
University College London, Dorking, RH5 6NT, UK}

\author[0000-0002-8627-4904]{Thomas J. Bogdan}
\affiliation{National Solar Observatory, 3665 Innovation Drive, Boulder CO 80303}
 
\author[0000-0001-6990-513X]{Roberto Casini}
\affiliation{HAO, National Center for Atmospheric Research, P.O. Box 3000, Boulder CO 80307-3000, USA}

\author[0000-0002-5084-4661]{Alfred G. de Wijn}
\affiliation{HAO, National Center for Atmospheric Research, P.O. Box 3000, Boulder CO 80307-3000, USA}

\author[0000-0001-5174-0568]{Philip G. Judge}
\affiliation{HAO, National Center for Atmospheric Research, P.O. Box 3000, Boulder CO 80307-3000, USA}

\begin{abstract}

The Visible Spectro-Polarimeter (ViSP) of the NSF Daniel K. Inouye Solar Telescope (DKIST) collected its Science
Verification data on May 7-8, 2021. The instrument observed multiple layers of a
sunspot atmosphere simultaneously, in passbands of \ion{Ca}{2} 397\,nm (H-line), \ion{Fe}{1} 630\,nm, and \ion{Ca}{2} 854\,nm, scanning the region with a spatial
sampling of 0.041\arcsec\ and average temporal cadence of 7.76 seconds, 
for an 38.8 minute duration. The slit moves southward across the 
plane-of-the-sky at 3.83 km/s.
The spectropolarimetric
scans exhibit prominent oscillatory `ridge' structures
which lie nearly perpendicular to the direction of slit motion (north to south). These ridges are visible in
maps of line intensity, central wavelength, line width, and both linear and circular polarizations. Contemporaneous Atmospheric Imaging Assembly observations indicate
these ridges are purely temporal in character and likely attributed to the
familiar chromospheric 3-minute 
umbral oscillations.

We observe in detail a steady umbral flash near the center
of the sunspot umbra. 
Although bad seeing limited the spatial resolution, the unique high signal-to-noise data enable us to
 estimate 
the shock
Mach numbers (${\approx}\,2$), propagation speeds (${\approx}\,9$ km/s), and their impact on 
the longitudinal magnetic field ($\Delta B \approx 50$ G), gas pressure, and temperature ($\Delta T/T \approx 0.1$) of the subshocks over 30 seconds. We also find evidence for rarefaction waves situated between neighboring
wave-train shocks. The \ion{Ca}{2} 854\,nm line width is fairly steady throughout the umbral flash except for a sharp 1.5 km/s dip immediately before, and comparable spike immediately after,
the passage of the shock front. This zig-zag in line width is centered on the subshock and
extends over 0.4\arcsec . 

\end{abstract}


\section{Introduction}

Well over half a century since their discovery by \cite{Beckers+Tallant1969}, \cite{Schultz+White1974}, our interest in umbral flashes continues unabated. They
remain one of the most striking dynamical chromospheric phenomena with roots that extend
deep into the umbral photosphere and possibly as far as the subsurface
magnetoconvection. They span numerous density scale heights and couple distinct
atmospheric layers. They provide a means to deliver mechanical energy to the 
optically-thin chromosphere, transition region, and lower corona. The extensive
monographs and reviews by \cite{Thomas+Weiss2012}, \cite{Weiss+Proctor2014}, and
\cite{Khomenko+Collados2015} provide an exhaustive summary and analysis of the 
research on umbral flashes through the first decade of the 21st Century. For more
recent efforts, from both observational and theoretical perspectives, one should 
consult, e.g., \cite{Madsenetal2015}, \cite{Thesis}, \cite{Songetal2017},
\cite{Kuzma2017}, \cite{Kuzmaetal2017}, \cite{Felipeetal2018}, 
\cite{Houstonetal2018}, \cite{Stangalinietal2018}, \cite{Joshi+2018},
\cite{Ananetal2019}, \cite{Boseetal2019},
\cite{Henriquesetal2020}, \cite{Houstonetal2020}, \cite{Yurch2020},
\cite{Stangalinietal2021a}, \cite{Stangalinietal2021b},
\cite{Felipe2021}, \cite{Sadykovetal2021}, 
\cite{Snow+Hillier2021}, and \cite{Molnaretal2021}. The following earlier papers
are valuable and particularly germane to what follows: \cite{Lites1992}, \cite{Centenoetal2006}, \cite{Pietarilaetal}, \cite{Pietarila2007},
\cite{Centenoetal2009},
\cite{Bard+Carlsoon2010}, and \cite{Felipeetal2014}.

The present contribution to the subject adds to this body of knowledge by
providing a unique high-spatial, spectral, and temporal resolution glimpse of
ten minutes in the hour or two lifetime of a mature umbral flash. This is achieved
by employing the inaugural Science Verification (SV) data from the Visible Spectro-Polarimeter \citep[ViSP,][]{DeWijnetal2022} attached to the NSF Daniel K. Inouye Solar Telescope \citep[DKIST,][]{2020SoPh..295..172R}.

The ViSP is one of the first-light instruments of the DKIST.
Here, we analyze spectropolarimetric data that were obtained by the ViSP during its SV campaign on May 8, 2021.
The ViSP uses
three spectral arms and cameras to measure the full state of polarization (Stokes $I$, $Q$, $U$, $V$) simultaneously
over three different spectral windows. For SV, the instrument was configured to observe simultaneously in the passbands of the photospheric lines of \ion{Fe}{1} around 630\,nm, and the two chromospheric resonance lines of \ion{Ca}{2} at 397\,nm (the H-line) and 854\,nm (2nd line of the IR triplet). 
The telescope was pointed to the
northernmost sunspot of AR\,12822 (the only active region on the solar
disk at the time), near the NE limb of the Sun. The line-of-sight (LOS) is
inclined 63$^\circ$ ($\mu
\approx 0.45$) 
from sunspot's zenith, and is tilted 23$^\circ$ off the E-W direction.

\section{Observations}

\begin{figure*}
\centering \includegraphics[width=16cm]{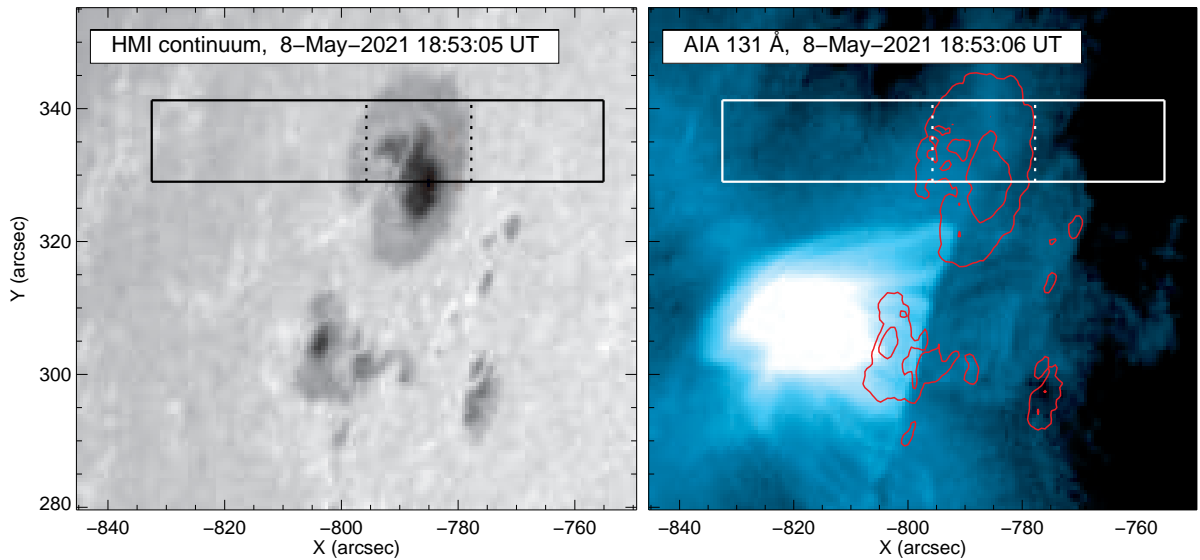}
\caption{ Left: HMI continuum showing AR 12822. The solid black rectangle marks
  the ViSP FOV. The inset dashed box shows the cropped FOV employed in
  Figure \ref{fig:images}. Right: AIA 131
  \AA\ observations showing the solar flare in progress. As in the left panel, the
  white/dashed rectangles show the full/cropped ViSP FOV. Red contours mark
  the sunspot location, as seen in HMI continuum (left panel). The bright feature in the right panel shows the flare loops of a C-class flare peaking prior to the beginning of the ViSP observations. }
\label{fig:context}
\end{figure*}

The ViSP polarimetric scan ran
from 18:52:39 to 19:31:19 UT 8 May 2021. It consists of 300 contiguous slit positions, separated by the slit width of 0.041\arcsec. It covers an angular extent of 12.3\arcsec\ in the N-to-S direction. The field-of-view (FOV) covered by the lowest-magnification  spectral arm of the ViSP (vcc1)
is shown in Figure \ref{fig:context}. The extent of the FOV along the slit, and the corresponding spatial sampling by the detector, are different for the three spectral arms: 75.9\arcsec\ with  0.030\arcsec/px for camera vcc1 (\ion{Fe}{1}), 62.3\arcsec\ with 0.024\arcsec/px for camera vcc2 (\ion{Ca}{2} H-line), and 49.4\arcsec\  with 0.019\arcsec/px for camera vcc3 (\ion{Ca}{2} 854\,nm). 

\begin{figure*}
\centering \includegraphics[width=14cm]{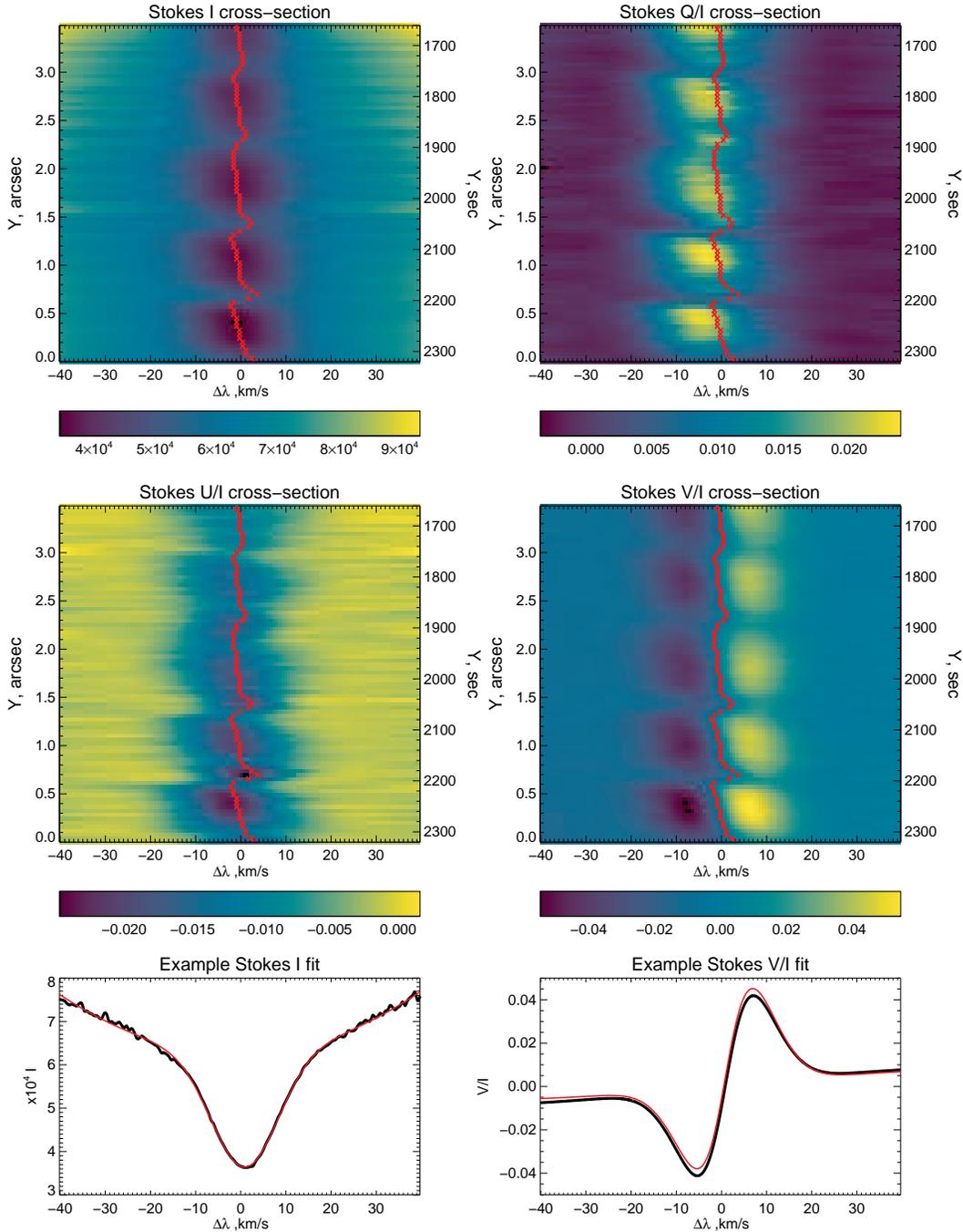}
\caption{Full Ca II 854.2 nm spectral data of the fractional stokes parameters observed at the $X$ position and along the $Y$ range corresponding to the magenta dashed line in Figure \ref{fig:images}. The red line tracks the center-of-gravity of Stokes $I$. The bottom panels show an example fit (red) to Stokes I and V/I data (black), for the spectra at Y=2150 sec in the top panels.}
\label{fig:stokes}
\end{figure*}

The dwell time for one slit position was 7.58 sec, consisting of 25~modulation cycles of 10~states each at a camera frame rate of 33~Hz. Because of the different intensity levels and instrument throughput values of the three spectral channels, the duty cycles of the three cameras were dramatically different, with the \ion{Fe}{1} spectral window being integrated only 20\% of the time with 6~ms exposures. The two \ion{Ca}{2} spectral lines were observed with 66\% duty cycle (20~ms exposures), to build sufficient signal-to-noise in the chromospheric Stokes $Q$, $U$, and $V$ signals. After integration, the slit was advanced by one full slit width (0.041\arcsec) for a new integration. Each of the 300 steps take 0.18~s, resulting in a time interval between slit positions of 7.76~s, for a total raster duration of 38.8 minutes.
For the observations we analyze here, the ViSP FOV was rastered over 300 steps, with a total scan time of 38.8 min. This corresponds to an 
average slit speed projected onto
the plane-of-the-sky (POS) of $v_s = 3.83~\mathrm{km}/\mathrm{s}$.

The FOV covered by the ViSP scan is marked in Figure \ref{fig:context} relative to the Helioseismic and Magnetic Imager \citep[HMI, ][]{Schou2012} white-light continuum and the Atmospheric Imaging Assembly \citep[AIA, ][]{Lemen2012} extreme-ultraviolet 131 \AA\ images. 
Shortly before the start of the ViSP data acquisition, a C8.6-class solar flare 
occurred to the SE of the active region. The location of the flare loops, 
which peaked in emission at 18:45 UT, are the bright feature in the right panel of Figure \ref{fig:context}.
The
northern-most foot point of the flare loops intersects the penumbra of
the larger sunspot observed by ViSP.
Unfortunately, the short temporal duration and spatial coverage of the scan
make it impossible to distinguish between 
flare-induced sunquakes and ubiquitous sub-photospheric umbral wave sources in AR 12822---e.g., \cite{Kosovichev+Sekii2007}, \cite{Ionescuetal2017},
\cite{Millaretal2021}. A cursory examination of the HMI and AIA contemporaneous data also
fails to identify any obvious oscillatory variations that can 
be unequivocally attributed
to the flare. 

For the analysis that follows, we restrict our attention to the cropped portion of
the FOV that lies above the sunspot where strong magnetic fields and sensible
polarization signal should be present. The two vertical dashed lines near
$X = -795$ and $X=-775$ mark the E-W 
boundaries of this region-of-interest
(ROI).

\subsection{Data Reduction}

Figure \ref{fig:stokes} displays a typical spectropolarimetric data 
cross-section obtained with
the vcc3 camera. A robust oscillatory signal is present in all four Stokes parameters.
The intensity profile consists of a narrow Gaussian core ($\Delta \lambda \approx 20$ km/s)
and broad Lorentzian wings.

During the 8 May 2021 SV
observations, a slight defocus of the ViSP collimator caused the appearance of considerable astigmatism for spectral configurations significantly away from Littrow. While vcc1 practically did not suffer any spectral defocus because of its proximity to the Littrow configuration ($3.31^\circ$), the spectral resolution of cameras vcc2 and vcc3 (respectively $7.52^\circ$ and $20.07^\circ$ from Littrow) was significantly impacted.
Therefore, we did not attempt a full inversion of the \ion{Ca}{2} 854\,nm Stokes profiles. Instead, we simply 
applied a single Gaussian fit to the core and near wings of the intensity and $I\pm V$ profiles to estimate the line core intensity, central wavelength, line width, and Zeeman splitting. A representative Gaussian fit is presented in Figure \ref{fig:stokes}. Assuming the line centers of Stokes $I$, $Q$ and
$U$ occur close to the same wavelength position, the Gaussian fit also
provides the position and value of the peak Stokes
$Q$ and $U$ emissions. This assumption holds for the majority of the FOV, but has limitations discussed in section \ref{summary}.
From these linear polarization measurements, we
calculate the linear polarization degree and azimuth,
\begin{eqnarray}
P &=& \frac{\sqrt{Q^2 + U^2}}{I}\;,
  \label{eqn:P}  \\
  \Phi &=& \frac{1}{2}\arctan{\frac{U}{Q}}\;.  \label{eqn:azi}
\end{eqnarray}
When the linear polarization is dominated by the Zeeman effect, the azimuth $\Phi$ also gives the direction of the magnetic field projected onto the POS (with a $180^\circ$ ambiguity).
The circularly polarized Stokes $V$ signals
yield the longitudinal magnetic field strength $B_{\rm LOS} \equiv B\cos\Theta$, where $\Theta$ is the angle between the magnetic field vector and the
LOS. A simple application of the weak-field approximation \citep[e.g.,][]{Landi+Landolfi2004, Centeno2018} appears to be adequate to provide good profile fits down to the noise level of the polarized signal (about $0.2$\% of Stokes $I$, in these observations), and was therefore used to estimate $B_{\rm LOS}$.

A proper description of the atmospheric dynamics at the time of the ViSP observations requires an adequate wavelength calibration of the spectral data. Because of the visibility of many telluric (i.e., Earth atmosphere's) molecular lines in the \ion{Fe}{1} 630\,nm spectral range, this is rather easily achieved for the vcc1 data. In contrast, the lack of strong telluric lines in the \ion{Ca}{2} 854\,nm spectrum makes a precise wavelength calibration of the vcc3 data more difficult. The significant spectral smearing of these data due to the ViSP astigmatism further complicates the matter. In order to arrive at a meaningful estimation of the chromospheric dynamics in the ViSP data, we followed the following procedure:
\begin{enumerate}
    \item We identified a small ($\sim3$\arcsec$\times$4\arcsec) QS region to the E of the sunspot, where the photospheric dynamics is  dominated by the granulation, resulting in highly symmetric distributions of positive and negative Doppler shift amplitudes for the lines of \ion{Fe}{1} at 629.8\,nm and 630.1\,nm, with practically zero mean LOS velocity over the region. Referencing both lines to the O2 telluric line at 629.8\,nm, we estimated a photospheric blue-shift of 1.65\,km/s, consistent with the solar rotational velocity expected at the  heliocentric coordinates of the observations;
    \item We determined the center-of-gravity of the \ion{Fe}{1} 853.8\,nm line in the vcc3 data, and assumed it to be at rest with the photosphere probed by the \ion{Fe}{1} lines in vcc1;
    \item Finally, we referenced the center-of-gravity of the chromospheric \ion{Ca}{2} 854.2\,nm line to the photospheric reference of the \ion{Fe}{1} 853.8\,nm line.
\end{enumerate}
The result of this wavelength calibration is that the zero-velocity reference of the \ion{Ca}{2} 854.2\,nm Doppler map in Fig.~3, averaged over the entire map, is red-shifted by about 1.53\,km/s with respect to the QS photosphere.

Recent work by \citet{Felipe2021}, suggests that
magnetic field oscillations should not easily be detected with the \ion{Ca}{2}
854\,nm line. However, because of the sunspot's
location close to the limb, the $B_{\rm LOS}$ measurements derived from Stokes
$V$ need not \textit{necessarily} relate to changes in
$|\bm{B}|$, but may well contain fluctuations in field orientation.

\begin{figure*}
\centering \includegraphics[width=15.5cm]{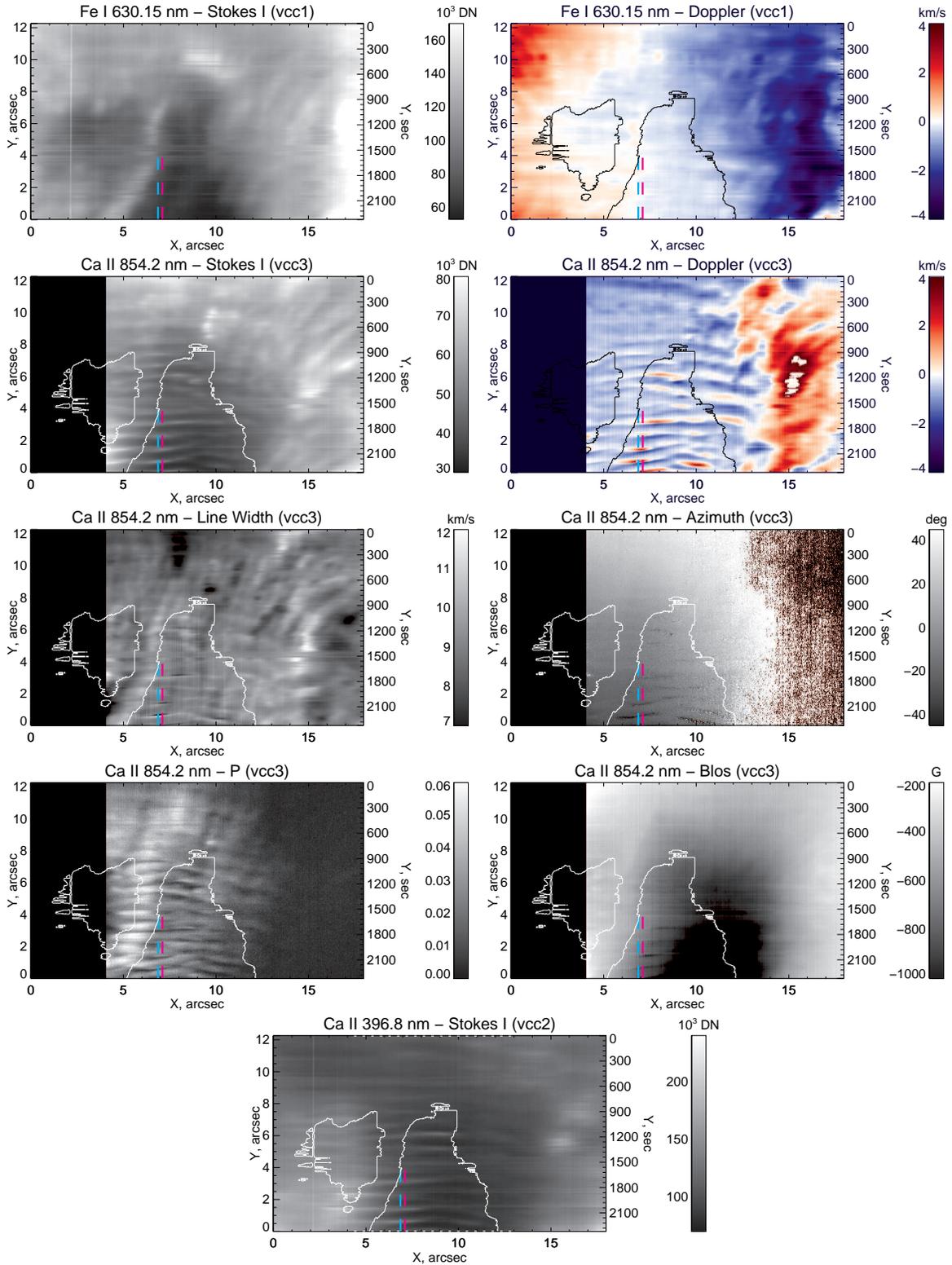}
\caption{ Active region scans from the ViSP SV
campaign. The plots show a cropped subset of the full FOV, as
  shown in Figure \ref{fig:context}. Each image is labeled with the
  spectral line and observed parameter. The cyan and magenta dashed lines shows the
  location of the umbral flash cross-sections examined in Figure
  \ref{fig:zoom}. 
 The solid white/black contour indicates the location of the sunspot umbra as determined from \ion{Fe}{1} 630\,nm intensity (top-left panel). We omit Ca II 854.2 data below 4\arcsec\ due to artifacts of poor fits in this region.}
\label{fig:images}
\end{figure*}  

Figure \ref{fig:images} shows data from the ViSP polarization scan over
the sunspot ROI indicated in Figure \ref{fig:context}. The
top row displays intensity in the core of the photospheric \ion{Fe}{1}
630\,nm line and its LOS Doppler velocity. The second row
provides the same for the chromospheric \ion{Ca}{2} 854\,nm line; 
derived magnetic parameters are presented in rows two to four. 
Finally, \ion{Ca}{2} H 397\,nm intensity is presented in the bottom panel.

The thin black/white contours indicate the approximate location of the umbra-penumbra boundary determined from the \ion{Fe}{1} 630\,nm line intensity. This is the boundary
between the umbra and the penumbra at the {\em photosphere}. The \ion{Ca}{2} resonance lines
are expected to form 
700 - 1400 km above this line.
Owing to the large inclination of the LOS, the projection on the POS will 
shift the position of the \ion{Ca}{2}
signal directly above the \ion{Fe}{1} signal some one-half to one arcsecond to the ENE. Wavelength-dependent atmospheric refraction will also shift
the chromospheric and photospheric signals on the POS. A light bridge separates the
E satellite umbra from the larger central umbra. Present in the E (left) umbra is the vertical ViSP hairline - used for calibration between the instrument cameras.

In Figure \ref{fig:images}, the rastered ViSP data obtained at each
value of $X$ are plotted in $Y$ against both slit position (left axis) and the time of
the data acquisition (right axis). Convenient origins in time and N-S slit
position have been selected to facilitate the analysis and display. We use both coordinates
interchangeably in what follows to emphasize that the ViSP scan via a 
stepped slit necessarily mixes N-S spatial and temporal variations.

\subsection{Summary of the Scans}\label{summary}

The sunspot structure is readily apparent in the photospheric \ion{Fe}{1} 630\,nm intensity plot, showing the larger central and smaller E 
satellite (left) umbrae (also seen in HMI, Figure \ref{fig:context}), separated by a thin light bridge. The surrounding
penumbra is visible to the N and W (right) of these regions.
There is no quiet sun in our ROI.
The Doppler velocity scans exhibit the expected Evershed flow in the
photosphere and the reverse chromospheric inflow from above
\citep[e.g.,][]{Teriacaetal2008}.

Prominent in the \ion{Ca}{2} scans are a series of nested ridges, 
running almost parallel to the
horizontal E-W direction across the image. These ridges are striking in
the Doppler velocity and $P$ scans. 
They are also present in line width, azimuth, $B_{\rm LOS}$,
and intensity. We see these same ridges in the H-line intensity. 
The average N-S separation of these ridges in $Y$ is on the
order of $0.5\arcsec$. A second more widely-separated set of horizontal ridges is also discernible in
the \ion{Fe}{1} scans. Their N-S spacing is slightly in excess of $1\arcsec$. Both sets of ridges are not strictly horizontal, but they
are well-resolved over 300 pixels from N to S, and highly-structured---exhibiting both curvature and
variation in brightness/thickness from E to W.

Observations of similar oscillatory ridge-like structures, sometimes referred to as 'herringbone' patterns, were first observed in the photosphere by works including \citet{Thomas1984} and \citet{lites1998}, and later in the chromosphere with the Dunn Solar Telescope / Horizontal Grading Spectrograph by \citep{Balasubramaniam2008}.

The moving slit yields an image scan that may be neither purely
temporal nor spatial, but could be a mixture of the two. 
It is not possible to determine from these observations alone whether the 
ridges originate from a large-scale 
coherent temporal
pulsation of the entire umbra, or from the moving slit 
passing over a static spatial 
undulation, or from a sinusoidal disturbance progressing northward or southward
across the sunspot. 
Previous observations of 'herringbone' ridges are captured by sit-and-stare measurements, which produce images of x against t. Because ViSP's moving slit creates images convolving spatial and temporal information in the vertical axis (creating an image of x against y and t), we lay out an argument in this section that the ridges observed by ViSP are not spatially-resolved structures, but are temporally-resolved, similar to the previous sit-and-stare observations.

Consider a progressive oscillatory disturbance moving northward across the POS at a speed $c$
with a spatial wavelength $\lambda_c$, and a temporal period $T_c$.
The {\em relative} velocity between the average southward speed of the stepped moving slit 
and the progressive disturbance determines the resulting separation of ridges in $Y$. 
The relationship between the true 
(subscript ``$c$") and
observed (no subscript) 
wavelengths and periods ($\lambda_c$, $\lambda$ and $T_c$,
$T$ respectively), are
\begin{equation}
  \frac{v_s + c}{c} = \frac{T_c}{T} =
  \frac{\lambda_c}{\lambda}\frac{v_s}{c}, \ 
  \label{eqn:doppler1}
\end{equation}
\begin{equation}
  c = \frac{v_s}{T_c/T-1} = v_s \biggl( \frac{\lambda_c}{\lambda}-1
  \biggr)\;.
  \label{eqn:doppler2}
\end{equation}
The observed period $T$ and wavelength $\lambda$ are related by $\lambda = T v_s$,
where $\lambda$ is the physical
distance between ridges on the ViSP scans, and $T$ is the time
between observations of the ridges. For oscillatory disturbances progressing
southward, one simply replaces $c$ by $-c$ in these expressions.

Notice that in the extreme case of a coherent spatial pulsation of the entire
sunspot ($\lambda_c,c\to\infty$)  the moving slit records the true period
$T = T_c$. Likewise for a static spatial undulation ($T_c = \infty$), the 
moving slit returns the true wavelength $\lambda = \lambda_c$. Between these
two extremes neither the recorded wavelength nor period will match their true
POS values.

To distinguish spatial from temporal behavior
we compare the ViSP observations 
with contextual (contemporaneous) observations of the photosphere/chromosphere 
obtained by HMI (intensity and Doppler measurements) and AIA (304 \AA, 1600 \AA\ and 1700 \AA). These data have poorer spatial
and temporal (per slit-position) resolution, however they 
have the advantages that they do not mix the spatial and temporal information and
they extend (in both space and time) well beyond the ViSP scan. These data
indicate fixed, spatially-coherent, patches of brightness oscillations in the sunspot umbra.
This is entirely consistent with the usual behavior of chromospheric 3-minute
umbral oscillations \citep{Khomenko+Collados2015}. 
Because the period of the AIA brightness oscillations match the period of the oscillations observed by ViSP, the ViSP ridges are not Doppler shifted by the moving slit. This reveals that the ridges do not arise from spatially-resolved waves traveling across the POS, but are instead
purely temporal in character as the ViSP slit moves over the pulsating region.

Because of their spatially-coherent temporal and periodic nature, the ridges observed by ViSP \textit{cannot} be a result of fine spatial structures within the umbra, such as intermixing hot and cool plasma elements \citep{Socas-Navarro2000,Socas-Navarro2009}, filamentary structure \citep{Socas-Navarro2009}, two-component umbral structure \citep{Centeno2005,delacruz2013}, small-scale dynamic fibrils \citep{Rouppe2013}, cool jet-like structures \citep{Yurchyshyn2014}, etc.

In the penumbra to the NW of the umbra, the ridges are much broader and
have a significant tilt relative to the $X$ and $Y$ directions. Here our slit
is probably sampling some combination of temporal and spatial variability appropriate
to running penumbral waves. 
    
\section{The Umbral Flash}

\begin{figure*}
\centering \includegraphics[width=18cm]{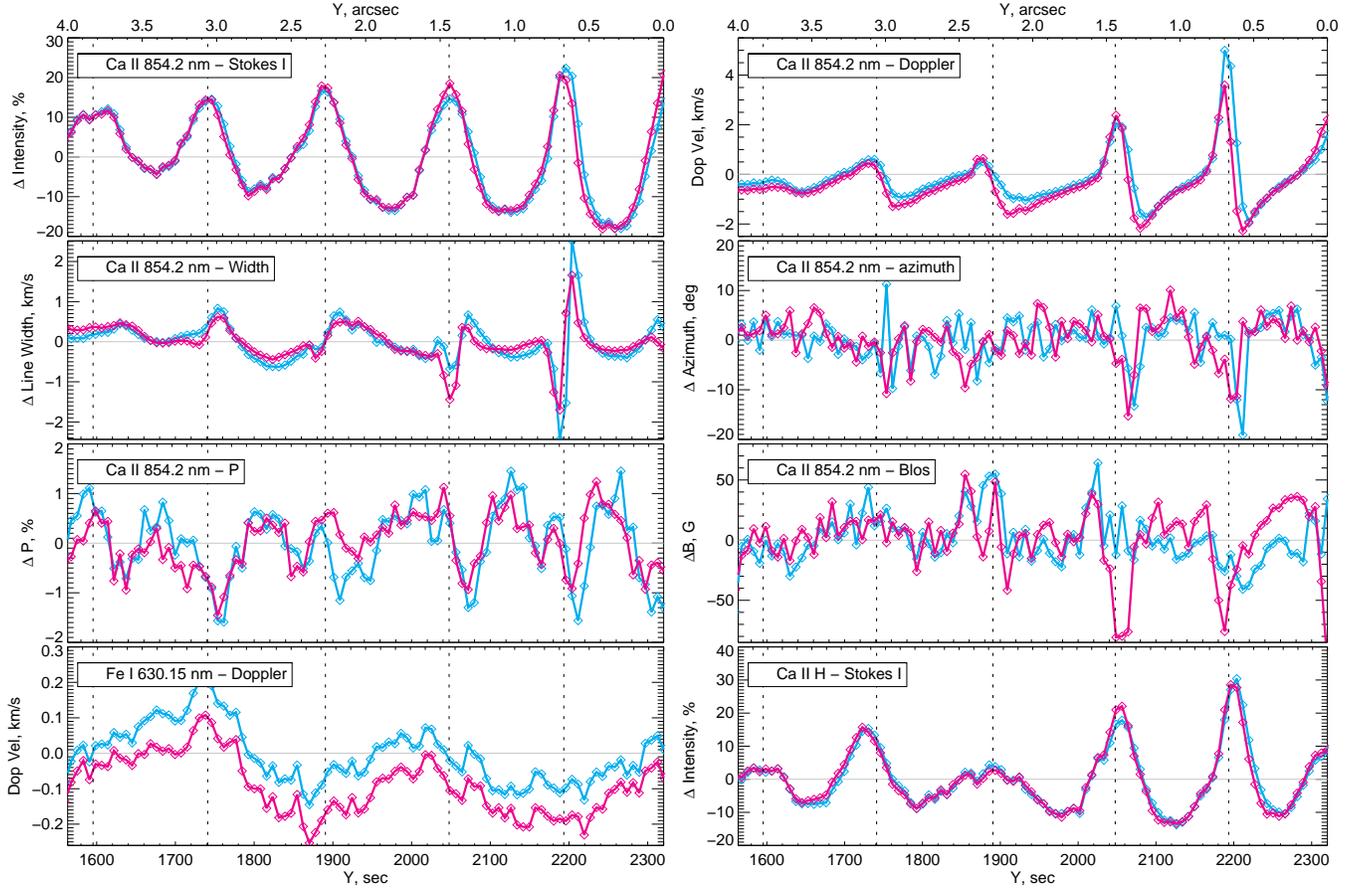}
\caption{Detrended intensity, line width, azimuth, P and B LOS cross-sections, and raw Doppler cross-sections, along the cyan and magenta dashed lines in Figure \ref{fig:images}. Vertical dashed lines show the approximate locations of the \ion{Ca}{2} 854\,nm intensity peaks. }
\label{fig:zoom}
\end{figure*}

Figure \ref{fig:zoom} provides a closer look at 
a portion of the chromospheric ridge structure. 
Here we plot a N-S cross-section through the \ion{Ca}{2} parameters along 
the dashed cyan and magenta line segments plotted in Figure \ref{fig:images}. 
The spectropolarimetric cross-section for the magenta line segment appear in Figure
\ref{fig:stokes}.
The lower axis 
of Figure \ref{fig:zoom} uses the temporal $Y$ coordinate for the
data acquisition and the upper axis uses the N-S spatial position of
the slit at the time the
data were obtained.
With the exception of Doppler velocity, the parameters have all been detrended by a polynomial fit to suppress
gradual variations across the ROI and to highlight the oscillatory signal. We also note that within this region, the line center of Stokes $Q$ differs from Stokes $I$ by about 3 km/s, as shown in Figure \ref{fig:stokes}. Line centroids of Stokes $U$ and $V$ still follow
$I$.

These data show a particularly revealing example of a chromospheric umbral flash, with the wave-train of shocks and rarefactions present in most of the \ion{Ca}{2} parameters.
For the chromospheric intensities, 
the extended shock compressions and intervening rarefaction
cross-sections are broad, well-resolved, and symmetric about peaks and valleys. 
The dashed vertical lines in Figure \ref{fig:zoom} show the approximate locations of the
shocks as identified by eye.
The \ion{Ca}{2} 854\, nm Doppler and line-width time series
differ significantly from the intensity profiles. Instead of symmetric rise and 
fall profiles, we see a steep rise and precipitous fall characteristic of a
thin embedded subshock. Indeed the
largest excursion near $Y=2200$ seconds occurs across 
just 4 $Y$-pixels/timesteps.
The line width
shows a somewhat broader zig-zag about the subshock.

In the magenta cross-section, the line-of-sight magnetic field (i.e., Stokes $V$) oscillates 180 degrees
out of phase
with the intensity parameters. The azimuth and linear polarization are 
also well-correlated with the subshock
but have lower signal-to-noise
ratios. There is a hint that they exhibit a negative
spike at the greatest positive excursion of the Doppler width (equivalently, 
negative excursion of the Doppler velocity). 

A careful examination of the contemporaneous AIA and HMI data indicates the umbral flash
is confined to $Y$ values less than 3 to 4 arcseconds, and it has been oscillating 
there with a slowly varying
temporal frequency throughout the duration of the ViSP data acquisition.
The increase in the amplitude of the oscillation from left to right in Figure \ref{fig:zoom} 
is consistent with the slit moving southward into the 
stronger core of this stationary, steady, umbral flash. The
temporal oscillation frequency present in Figure \ref{fig:zoom} is the same as that 
obtained from the AIA data during this epoch, meaning it has not been \textit{Doppler shifted} by the moving ViSP slit (Equation \ref{eqn:doppler2}). This is consistent with the umbral flash oscillating 
in place over a fixed spatial region near the center of the umbra. In other words, the
motion of the slit does not shift the observed frequency from the true 
oscillation frequency.
The spatial wavelength (bottom abscissa scale), however, is purely an artifact of
the sit-stare-step method, and
is given by $T_c/v_s$.

It is possible to estimate the parameters for these shocks. For example, 
let us consider the blue curve in the neighborhood of $Y=2200$. We find the peak POS 
pre-shock down-flow velocity
at $Y=2190$ to be 5.22 km/s. The post-shock up-flow velocity is 1.70 km/s. 
Dividing by
the cosine of the inclination angle (0.443) and assuming the motions are along a vertical 
magnetic flux tube near the center of the umbra we obtain a preshock downflow velocity of
$u_+ =$ 11.8 km/s and a postshock upflow velocity of 
$u_- =$ 3.83 km/s. 

\begin{figure*}
\centering \includegraphics[width=14cm]{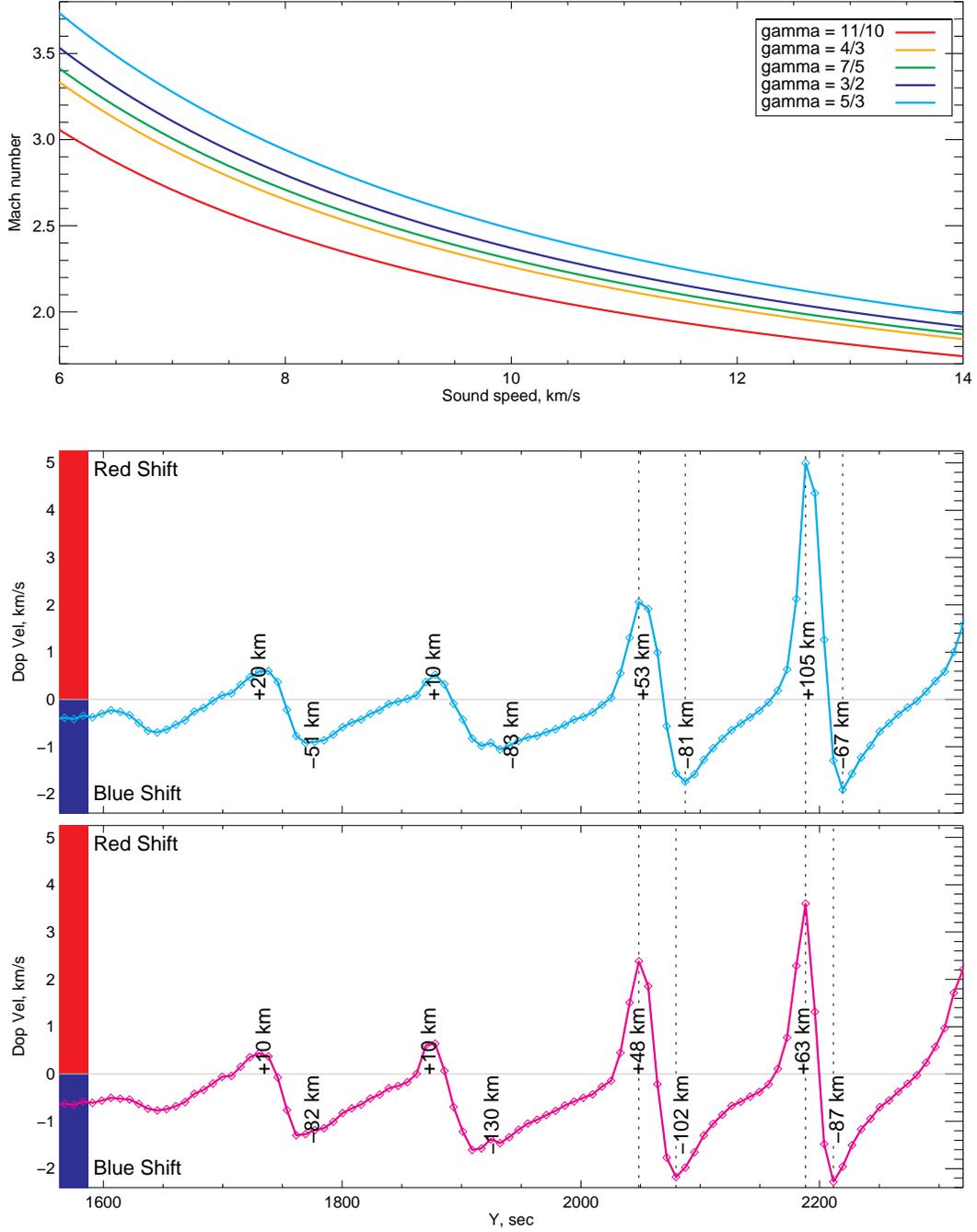}
\caption{
Top: Mach number of the largest shock in our observations (at Y$\approx2200$ seconds in the cyan cross-section), for a reasonable range of sound speed and $\gamma$ values.
Middle/Bottom:
Cross-sections of Ca II 854.2 nm Doppler velocity along the cyan (middle) and magenta (bottom) cross-sections in Figure \ref{fig:images} Dashed lines indicate the shock locations. Following time, one sees a near-constant
deceleration from blueshift to redshift behind the nonlinear pulses. As the shock forms, we see increased deceleration into red shift, followed by a rapid acceleration into blueshift, before repeating into the following shock.
}
\label{fig:mach}
\end{figure*}

Suppose the shock is moving upward 
along the vertical magnetic flux tube with a speed of $v$ km/s. In the rest frame
of the shock, the ratio of the upstream to downstream velocities (and downstream to 
upstream densities) is given by the
usual gas-dynamic Rankine-Hugoniot relations,
\begin{equation}
    {v+u_+ \over v-u_-} = {(\gamma +1)M^2 \over 2 + (\gamma-1)M^2} 
    = {\rho_- \over \rho_+}\;,
\end{equation}
where $M = (v+u_+)/c_s$ is the upstream Mach number, $c_s$ is the adiabatic sound speed,
and $\gamma$ is the ratio of specific heats. Setting $v+u_+ = c_sM$, and defining $\alpha$ as $\alpha = c_s/(u_++u_-)$,
we obtain the following expression for the Mach number:
\begin{equation}
    M = {(\gamma+1) \over 4\alpha} + \sqrt{ 1 + {(\gamma+1)^2 \over 16 \alpha^2}}\;.
\end{equation}
Once $M$ is obtained, the shock velocity $v$ is readily computed, as well as
the ratio of the downstream to upstream gas pressure
\begin{equation}
    {p_- \over p_+} = {2\gamma M^2 - (\gamma -1) \over \gamma +1}\;,
\end{equation}
and the analogous temperature ratio.

The ViSP observations give $u_+ + u_- =$ 15.6~km/s for the strongest blue shock 
shown in Figure \ref{fig:zoom}. 
Taking, for example, $\gamma = 1.1$ to account for Hydrogen ionization, and $c = 10~\mathrm{km}\,\mathrm{s}^{-1}$%
, one obtains: $M=2.11$, $v=9.3$~km/s,
$\rho_-/\rho_+=3.82$,
$p_-/p_+=4.62$,
and $T_-/T_+=1.21$.
Larger sound speeds produce weaker shocks which propagate faster; larger ratios
of specific heats produce stronger shocks which also propagate faster. The top panel of Figure \ref{fig:mach}
displays the Mach number (ordinate) for this strongest 
observed blue shock ($u_+ + u_- =$ 15.6~km/s) for
a range of upstream sound speeds ($c$, abscissa) and 
several adiabatic indices (colored curves). We note that these parameters are estimates, given that the Rankine-Hugoniot relation assumes an idealized infinite planar shock propagating along a constant uniform magnetic field. The fact we see a molderate decrease in magnetic field strength behind the shock does not invalidate these estimates.

The Stokes $V$ and linear polarization $P$ signals are broadly consistent with a 
modest expansion of the cross-sectional area of
the magnetic flux tube (decrease in magnetic field strength) expected from the
gas pressure increase behind the shock. The nominal magnetic field strength
is of order 500-600 G, so $\Delta B/B \approx 0.1$. As would be expected from such
an interpretation, the magnetic fluctuations are detected only for the strongest
shocks. 

The variation in the line-width zig-zag across the shock is interesting and robust.
It too is most pronounced for the strong shocks, and is
largely absent in the nonlinear pulses. We are
presently unable to identify a compelling physical explanation for it. The
dramatic change in slope of the Doppler velocity in passing from blue-shifted
(negative values) to red-shifted (positive values) suggests a 
quasi-free-fall rarefaction front may
lie in between shocks. The attendant rapid expansion of the material could lead
to cooling and might also tend to stretch out, or elongate turbulent motions set
up in the post-shock flow. Both of these effects {\em could} contribute to a reduced
line width. Likewise the enhancement of the line-width downstream of the shock
could arise from a combination of compressive heating and post-shock turbulence.
On the other hand, one must acknowledge that the line source function may encompass
non-equilibrium and dynamical complications, including, for example, analogous K2V
and K2R emission processes observed in the K-line. 
A definitive explanation of this line width zig-zag 
will require a careful spectral
synthesis.

The lack of asymmetry in Stokes $I$ indicates the
radiative transfer in the line core is optically-thick around the shock compression.
The ratio of the maximum to minimum values of Stokes $I$ is close to 3/2---the fourth
root of this number is 1.10, which is comparable to, but comfortably below our estimate
of $T_-/T_+$ of 1.21 given above. 

One can compute the integral under the undetrended Doppler velocity to determine
if there is a net flow of material along the magnetic flux tube over one cycle of
oscillation. The areas (expressed in km) between the zero-crossings of the blue and
magenta curve POS Doppler velocities are provided in the bottom panels of Figure \ref{fig:mach}. We have
not divided by the cosine of the inclination angle to the LOS (0.443). Taking into 
account this factor of 2.25 one observes that the range of upward and downward
displacements per oscillation cycle are no more than a density scale height. 
For the weaker nonlinear pulses there is a net upward motion of the material per cycle.
However, for the strongest shock the net motion is downward. The former behavior
is consistent with the general notion
that nonlinear waves contribute an upward-directed pressure
gradient and tend to elevate material in a flux tube relative to its 
neutral hydrostatic 
altitude. In the latter case, where strong localized shocks are present, the
precipitous quasi-free-fall redshift of the material in front of the shock may
carry
the material farther than the subsequent upflow behind the shock. Indeed, it is 
tempting to speculate that the over time, the nonlinear pulses eventually
raise too much
plasma above its neutral level and the subsequent formation of strong shocks
in the quasi-free-fall downflow is how the mature umbral flash addresses this
untenable situation.

The complex projection and atmospheric
refractive POS offsets between the chromospheric and photospheric
oscillations make it difficult
to investigate any potential photospheric roots of the
umbral flash. Never-the-less the \ion{Fe}{1} Doppler velocity appears to attain
its maximal blue shifts and red shifts close to the time of passage
of the chromospheric shocks. Equivalently,
the underlying photospheric oscillation has a period 
close to 5 minutes, or nearly twice that of the 150 second separation
of shocks in the chromosphere. This is reminiscent of the period/frequency
doubling often seen in highly nonlinear systems.

\section{Conclusion}

In this paper we examined DKIST/ViSP Science Verification (SV) observations  to assess high-spatial and temporal spectropolarimetric observations of chromospheric and photospheric oscillations
in a sunspot. 
We detected the ubiquitous running
penumbral waves, chromospheric 3-min oscillations, and an umbral flash, with the detected periods affected by the motion of the moving ViSP slit. All of
these phenomena have been observed and studied for decades. However, included in these oscillations data are the Stokes $V$ and linear polarization ($ P = \sqrt{U^2 + Q^2}/I$) of the chromospheric Ca II 854.2 nm line, which provide
rich spectropolarimetric signatures of the umbral flash. 

Owing to residual spectrograph astigmatism at the time of the SV campaign, the
data did not achieve sufficient spectral resolution to justify a detailed Stokes inversion of \ion{Ca}{2} 854 nm. Despite these drawbacks, the SV observations provide tantalizing hints of previously unknown spatial, temporal and spectral behaviors associated with dynamical processes associated with a sunspot umbra. In particular, we find polarimetric evidence of a wave-train of shocks and rarefactions over time scales of 0.16 seconds, the likes of which, to the authors' best knowledge, have not been detected before. Across the shock train, we find values of $M=1.88$, $v=10.6~\mathrm{km}\,\mathrm{s}^{-1}$,
$\rho_-/\rho_+=3.15$,
$p_-/p_+=3.66$,
and $T_-/T_+=1.16$.
These parameters are consist with the ranges of shock parameters calculated in other work, e.g. \citet{Anan2019}.

As ViSP has begun science commissioning and
normal operations in 2022, new observations will further push the spatial and
spectral limitations of the data reported here, providing further clarity to the dynamic oscillations found within the chromosphere and photosphere.

\begin{acknowledgments}
\section*{Acknowledgments}
We are indebted to C. Beck and A. Eigenbrot (both NSO/DKIST) for their critical role in the development, testing, and improvement of the ViSP data calibration pipeline that produced the set of science observations analyzed here. We also thank all the DKIST and NSO teams that were instrumental to the acquisition of this dataset, namely, the Telescope Control, Polarization Analysis and Calibration, and Data Center teams, and the Wavefront Corrector instrument team, who were responsible for locking on the target during the ViSP observing run. RJF thanks support from STFC PhD studentship, NCAR Newkirk Fellowship, and Brinson Prize Fellowship. TJB gratefully acknowledges the grant of a Person of Interest appointment by the National Solar Observatory. The research reported herein is based in part on data collected with the \textit{Daniel K. Inouye Solar Telescope} (DKIST), a facility of the National Solar Observatory (NSO). NSO is managed by the Association of Universities for Research in Astronomy, Inc., and is funded by the National Science Foundation. Any opinions, ﬁndings and conclusions or recommendations expressed in this publication are those of the authors and do not necessarily reﬂect the views of the National Science Foundation or the Association of Universities for Research in Astronomy, Inc. DKIST is located on land of spiritual and cultural signiﬁcance to Native Hawaiian people. The use of this important site to further scientiﬁc knowledge is done so with appreciation and respect. This research has made use of NASA's Astrophysics Data System. This material is based upon work supported by the National Center for Atmospheric Research, which is a major facility sponsored by the National Science Foundation under Cooperative Agreement No.~1852977.

\end{acknowledgments}

\bibliography{bibliography}{} \bibliographystyle{aasjournal}

\begin{thebibliography}{}
\expandafter\ifx\csname natexlab\endcsname\relax\def\natexlab#1{#1}\fi
\providecommand{\url}[1]{\href{#1}{#1}}
\providecommand{\dodoi}[1]{doi:~\href{http://doi.org/#1}{\nolinkurl{#1}}}
\providecommand{\doeprint}[1]{\href{http://ascl.net/#1}{\nolinkurl{http://ascl.net/#1}}}
\providecommand{\doarXiv}[1]{\href{https://arxiv.org/abs/#1}{\nolinkurl{https://arxiv.org/abs/#1}}}

\bibitem[{{Anan} {et~al.}(2019{\natexlab{a}}){Anan}, {Schad}, {Jaeggli}, \&
  {Tarr}}]{Ananetal2019}
{Anan}, T., {Schad}, T.~A., {Jaeggli}, S.~A., \& {Tarr}, L.~A.
  2019{\natexlab{a}}, \apj, 882, 161, \dodoi{10.3847/1538-4357/ab357f}

\bibitem[{{Anan} {et~al.}(2019{\natexlab{b}}){Anan}, {Schad}, {Jaeggli}, \&
  {Tarr}}]{Anan2019}
---. 2019{\natexlab{b}}, \apj, 882, 161, \dodoi{10.3847/1538-4357/ab357f}

\bibitem[{{Balasubramaniam} {et~al.}(2008){Balasubramaniam}, {Pevtsov}, \&
  {Olmschenk}}]{Balasubramaniam2008}
{Balasubramaniam}, K.~S., {Pevtsov}, A.~A., \& {Olmschenk}, S. 2008, in
  Astronomical Society of the Pacific Conference Series, Vol. 383, Subsurface
  and Atmospheric Influences on Solar Activity, ed. R.~{Howe}, R.~W. {Komm},
  K.~S. {Balasubramaniam}, \& G.~J.~D. {Petrie}, 279

\bibitem[{{Bard} \& {Carlsson}(2010)}]{Bard+Carlsoon2010}
{Bard}, S., \& {Carlsson}, M. 2010, \apj, 722, 888,
  \dodoi{10.1088/0004-637X/722/1/888}

\bibitem[{{Beckers} \& {Tallant}(1969)}]{Beckers+Tallant1969}
{Beckers}, J.~M., \& {Tallant}, P.~E. 1969, \solphys, 7, 351,
  \dodoi{10.1007/BF00146140}

\bibitem[{{Besliu-Ionescu} {et~al.}(2017){Besliu-Ionescu}, {Donea}, \&
  {Cally}}]{Ionescuetal2017}
{Besliu-Ionescu}, D., {Donea}, A., \& {Cally}, P. 2017, Sun and Geosphere, 12,
  59

\bibitem[{{Bose} {et~al.}(2019){Bose}, {Henriques}, {Rouppe van der Voort}, \&
  {Pereira}}]{Boseetal2019}
{Bose}, S., {Henriques}, V. M.~J., {Rouppe van der Voort}, L., \& {Pereira}, T.
  M.~D. 2019, \aap, 627, A46, \dodoi{10.1051/0004-6361/201935289}

\bibitem[{{Centeno}(2018)}]{Centeno2018}
{Centeno}, R. 2018, \apj, 866, 89, \dodoi{10.3847/1538-4357/aae087}

\bibitem[{{Centeno} {et~al.}(2006){Centeno}, {Collados}, \& {Trujillo
  Bueno}}]{Centenoetal2006}
{Centeno}, R., {Collados}, M., \& {Trujillo Bueno}, J. 2006, \apj, 640, 1153,
  \dodoi{10.1086/500185}

\bibitem[{{Centeno} {et~al.}(2009){Centeno}, {Collados}, \& {Trujillo
  Bueno}}]{Centenoetal2009}
---. 2009, \apj, 692, 1211, \dodoi{10.1088/0004-637X/692/2/1211}

\bibitem[{{Centeno} {et~al.}(2005){Centeno}, {Socas-Navarro}, {Collados}, \&
  {Trujillo Bueno}}]{Centeno2005}
{Centeno}, R., {Socas-Navarro}, H., {Collados}, M., \& {Trujillo Bueno}, J.
  2005, \apj, 635, 670, \dodoi{10.1086/497393}

\bibitem[{{de la Cruz Rodr{\'\i}guez} {et~al.}(2013){de la Cruz
  Rodr{\'\i}guez}, {Rouppe van der Voort}, {Socas-Navarro}, \& {van
  Noort}}]{delacruz2013}
{de la Cruz Rodr{\'\i}guez}, J., {Rouppe van der Voort}, L., {Socas-Navarro},
  H., \& {van Noort}, M. 2013, \aap, 556, A115,
  \dodoi{10.1051/0004-6361/201321629}

\bibitem[{{De Wijn} {et~al.}(2022){De Wijn}, {Casini}, {Carlile}, \&
  {etal}}]{DeWijnetal2022}
{De Wijn}, A.~G., {Casini}, R., {Carlile}, A., \& {etal}. 2022, Solar Physics,
  297, 22

\bibitem[{{Felipe} {et~al.}(2014){Felipe}, {Socas-Navarro}, \&
  {Khomenko}}]{Felipeetal2014}
{Felipe}, T., {Socas-Navarro}, H., \& {Khomenko}, E. 2014, \apj, 795, 9,
  \dodoi{10.1088/0004-637X/795/1/9}

\bibitem[{{Felipe} {et~al.}(2018){Felipe}, {Socas-Navarro}, \&
  {Przybylski}}]{Felipeetal2018}
{Felipe}, T., {Socas-Navarro}, H., \& {Przybylski}, D. 2018, \aap, 614, A73,
  \dodoi{10.1051/0004-6361/201732169}

\bibitem[{{Felipe} {et~al.}(2021){Felipe}, {Socas Navarro}, {Sangeetha}, \&
  {Milic}}]{Felipe2021}
{Felipe}, T., {Socas Navarro}, H., {Sangeetha}, C.~R., \& {Milic}, I. 2021,
  ApJ, 918, 47, \dodoi{10.3847/1538-4357/ac111c}

\bibitem[{{Henriques} {et~al.}(2020){Henriques}, {Nelson}, {Rouppe van der
  Voort}, \& {Mathioudakis}}]{Henriquesetal2020}
{Henriques}, V. M.~J., {Nelson}, C.~J., {Rouppe van der Voort}, L. H.~M., \&
  {Mathioudakis}, M. 2020, \aap, 642, A215, \dodoi{10.1051/0004-6361/202038538}

\bibitem[{{Houston} {et~al.}(2018){Houston}, {Jess}, {Asensio Ramos}, {Grant},
  {Beck}, {Norton}, \& {Krishna Prasad}}]{Houstonetal2018}
{Houston}, S.~J., {Jess}, D.~B., {Asensio Ramos}, A., {et~al.} 2018, \apj, 860,
  28, \dodoi{10.3847/1538-4357/aab366}

\bibitem[{{Houston} {et~al.}(2020){Houston}, {Jess}, {Keppens}, {Stangalini},
  {Keys}, {Grant}, {Jafarzadeh}, {McFetridge}, {Murabito}, {Ermolli}, \&
  {Giorgi}}]{Houstonetal2020}
{Houston}, S.~J., {Jess}, D.~B., {Keppens}, R., {et~al.} 2020, \apj, 892, 49,
  \dodoi{10.3847/1538-4357/ab7a90}

\bibitem[{{Joshi} \& {de la Cruz Rodr{\'\i}guez}(2018)}]{Joshi+2018}
{Joshi}, J., \& {de la Cruz Rodr{\'\i}guez}, J. 2018, \aap, 619, A63,
  \dodoi{10.1051/0004-6361/201832955}

\bibitem[{{Khomenko} \& {Collados}(2015)}]{Khomenko+Collados2015}
{Khomenko}, E., \& {Collados}, M. 2015, Living Reviews in Solar Physics, 12, 6

\bibitem[{{Kosovichev} \& {Sekii}(2007)}]{Kosovichev+Sekii2007}
{Kosovichev}, A.~G., \& {Sekii}, T. 2007, \apjl, 670, L147,
  \dodoi{10.1086/524298}

\bibitem[{{Ku{\'z}ma} {et~al.}(2017{\natexlab{a}}){Ku{\'z}ma}, {Murawski},
  {Kayshap}, {W{\'o}jcik}, {Srivastava}, \& {Dwivedi}}]{Kuzma2017}
{Ku{\'z}ma}, B., {Murawski}, K., {Kayshap}, P., {et~al.} 2017{\natexlab{a}},
  \apj, 849, 78, \dodoi{10.3847/1538-4357/aa8ea1}

\bibitem[{{Ku{\'z}ma} {et~al.}(2017{\natexlab{b}}){Ku{\'z}ma}, {Murawski},
  {Zaqarashvili}, {Konkol}, \& {Mignone}}]{Kuzmaetal2017}
{Ku{\'z}ma}, B., {Murawski}, K., {Zaqarashvili}, T.~V., {Konkol}, P., \&
  {Mignone}, A. 2017{\natexlab{b}}, \aap, 597, A133,
  \dodoi{10.1051/0004-6361/201628747}

\bibitem[{{Landi Degl'Innocenti} \& {Landolfi}(2004)}]{Landi+Landolfi2004}
{Landi Degl'Innocenti}, E., \& {Landolfi}, M. 2004, {Polarization in Spectral
  Lines}, Vol. 307, \dodoi{10.1007/978-1-4020-2415-3}

\bibitem[{{Lemen} {et~al.}(2012){Lemen}, {Title}, {Akin}, {Boerner}, {Chou},
  {Drake}, {Duncan}, {Edwards}, {Friedlaender}, {Heyman}, {Hurlburt}, {Katz},
  {Kushner}, {Levay}, {Lindgren}, {Mathur}, {McFeaters}, {Mitchell}, {Rehse},
  {Schrijver}, {Springer}, {Stern}, {Tarbell}, {Wuelser}, {Wolfson}, {Yanari},
  {Bookbinder}, {Cheimets}, {Caldwell}, {Deluca}, {Gates}, {Golub}, {Park},
  {Podgorski}, {Bush}, {Scherrer}, {Gummin}, {Smith}, {Auker}, {Jerram},
  {Pool}, {Soufli}, {Windt}, {Beardsley}, {Clapp}, {Lang}, \&
  {Waltham}}]{Lemen2012}
{Lemen}, J.~R., {Title}, A.~M., {Akin}, D.~J., {et~al.} 2012, \solphys, 275,
  17, \dodoi{10.1007/s11207-011-9776-8}

\bibitem[{{Lites}(1992)}]{Lites1992}
{Lites}, B.~W. 1992, in NATO Advanced Study Institute (ASI) Series C, Vol. 375,
  Sunspots. Theory and Observations, ed. J.~H. {Thomas} \& N.~O. {Weiss}, 261

\bibitem[{{Lites} {et~al.}(1998){Lites}, {Thomas}, {Bogdan}, \&
  {Cally}}]{lites1998}
{Lites}, B.~W., {Thomas}, J.~H., {Bogdan}, T.~J., \& {Cally}, P.~S. 1998, \apj,
  497, 464, \dodoi{10.1086/305451}

\bibitem[{{L{\"o}hner-B{\"o}ttcher}(2016)}]{Thesis}
{L{\"o}hner-B{\"o}ttcher}, J. 2016, PhD thesis, Albert Ludwigs University of
  Freiburg, Germany

\bibitem[{{Madsen} {et~al.}(2015){Madsen}, {Tian}, \&
  {DeLuca}}]{Madsenetal2015}
{Madsen}, C.~A., {Tian}, H., \& {DeLuca}, E.~E. 2015, \apj, 800, 129,
  \dodoi{10.1088/0004-637X/800/2/129}

\bibitem[{{Millar} {et~al.}(2021){Millar}, {Fletcher}, \&
  {Milligan}}]{Millaretal2021}
{Millar}, D. C.~L., {Fletcher}, L., \& {Milligan}, R.~O. 2021, \mnras, 503,
  2444, \dodoi{10.1093/mnras/stab642}

\bibitem[{{Molnar} {et~al.}(2021){Molnar}, {Reardon}, {Cranmer}, {Kowalski},
  {Chai}, \& {Gary}}]{Molnaretal2021}
{Molnar}, M.~E., {Reardon}, K.~P., {Cranmer}, S.~R., {et~al.} 2021, \apj, 920,
  125, \dodoi{10.3847/1538-4357/ac1515}

\bibitem[{{Pietarila} {et~al.}(2007){Pietarila}, {Socas-Navarro}, \&
  {Bogdan}}]{Pietarila2007}
{Pietarila}, A., {Socas-Navarro}, H., \& {Bogdan}, T. 2007, \apj, 663, 1386,
  \dodoi{10.1086/518714}

\bibitem[{{Pietarila} {et~al.}(2006){Pietarila}, {Socas-Navarro}, {Bogdan},
  {Carlsson}, \& {Stein}}]{Pietarilaetal}
{Pietarila}, A., {Socas-Navarro}, H., {Bogdan}, T., {Carlsson}, M., \& {Stein},
  R.~F. 2006, \apj, 640, 1142, \dodoi{10.1086/500240}

\bibitem[{{Rimmele} {et~al.}(2020){Rimmele}, {Warner}, {Keil}, {Goode},
  {Kn{\"o}lker}, {Kuhn}, {Rosner}, {McMullin}, {Casini}, {Lin}, {W{\"o}ger},
  {von der L{\"u}he}, {Tritschler}, {Davey}, {de Wijn}, {Elmore}, {Fehlmann},
  {Harrington}, {Jaeggli}, {Rast}, {Schad}, {Schmidt}, {Mathioudakis},
  {Mickey}, {Anan}, {Beck}, {Marshall}, {Jeffers}, {Oschmann}, {Beard},
  {Berst}, {Cowan}, {Craig}, {Cross}, {Cummings}, {Donnelly}, {de Vanssay},
  {Eigenbrot}, {Ferayorni}, {Foster}, {Galapon}, {Gedrites}, {Gonzales},
  {Goodrich}, {Gregory}, {Guzman}, {Guzzo}, {Hegwer}, {Hubbard}, {Hubbard},
  {Johansson}, {Johnson}, {Liang}, {Liang}, {McQuillen}, {Mayer}, {Newman},
  {Onodera}, {Phelps}, {Puentes}, {Richards}, {Rimmele}, {Sekulic}, {Shimko},
  {Simison}, {Smith}, {Starman}, {Sueoka}, {Summers}, {Szabo}, {Szabo},
  {Wampler}, {Williams}, \& {White}}]{2020SoPh..295..172R}
{Rimmele}, T.~R., {Warner}, M., {Keil}, S.~L., {et~al.} 2020, \solphys, 295,
  172, \dodoi{10.1007/s11207-020-01736-7}

\bibitem[{{Rouppe van der Voort} \& {de la Cruz
  Rodr{\'\i}guez}(2013)}]{Rouppe2013}
{Rouppe van der Voort}, L., \& {de la Cruz Rodr{\'\i}guez}, J. 2013, \apj, 776,
  56, \dodoi{10.1088/0004-637X/776/1/56}

\bibitem[{{Sadykov} {et~al.}(2021){Sadykov}, {Kitiashvili}, {Kosovichev}, \&
  {Wray}}]{Sadykovetal2021}
{Sadykov}, V.~M., {Kitiashvili}, I.~N., {Kosovichev}, A.~G., \& {Wray}, A.~A.
  2021, \apj, 909, 35, \dodoi{10.3847/1538-4357/abd9c7}

\bibitem[{{Schou} {et~al.}(2012){Schou}, {Scherrer}, {Bush}, {Wachter},
  {Couvidat}, {Rabello-Soares}, {Bogart}, {Hoeksema}, {Liu}, {Duvall}, {Akin},
  {Allard}, {Miles}, {Rairden}, {Shine}, {Tarbell}, {Title}, {Wolfson},
  {Elmore}, {Norton}, \& {Tomczyk}}]{Schou2012}
{Schou}, J., {Scherrer}, P.~H., {Bush}, R.~I., {et~al.} 2012, \solphys, 275,
  229, \dodoi{10.1007/s11207-011-9842-2}

\bibitem[{{Schultz} \& {White}(1974)}]{Schultz+White1974}
{Schultz}, R.~B., \& {White}, O.~R. 1974, \solphys, 35, 309,
  \dodoi{10.1007/BF00151951}

\bibitem[{{Snow} \& {Hillier}(2021)}]{Snow+Hillier2021}
{Snow}, B., \& {Hillier}, A. 2021, \mnras, 506, 1334,
  \dodoi{10.1093/mnras/stab1672}

\bibitem[{{Socas-Navarro} {et~al.}(2009){Socas-Navarro}, {McIntosh}, {Centeno},
  {de Wijn}, \& {Lites}}]{Socas-Navarro2009}
{Socas-Navarro}, H., {McIntosh}, S.~W., {Centeno}, R., {de Wijn}, A.~G., \&
  {Lites}, B.~W. 2009, \apj, 696, 1683, \dodoi{10.1088/0004-637X/696/2/1683}

\bibitem[{{Socas-Navarro} {et~al.}(2000){Socas-Navarro}, {Trujillo Bueno}, \&
  {Ruiz Cobo}}]{Socas-Navarro2000}
{Socas-Navarro}, H., {Trujillo Bueno}, J., \& {Ruiz Cobo}, B. 2000, Science,
  288, 1396, \dodoi{10.1126/science.288.5470.1396}

\bibitem[{{Song} {et~al.}(2017){Song}, {Chae}, {Yurchyshyn}, {Lim}, {Cho},
  {Yang}, {Cho}, \& {Kwak}}]{Songetal2017}
{Song}, D., {Chae}, J., {Yurchyshyn}, V., {et~al.} 2017, \apj, 835, 240,
  \dodoi{10.3847/1538-4357/835/2/240}

\bibitem[{{Stangalini} {et~al.}(2018){Stangalini}, {Jafarzadeh}, {Ermolli},
  {Erd{\'e}lyi}, {Jess}, {Keys}, {Giorgi}, {Murabito}, {Berrilli}, \& {Del
  Moro}}]{Stangalinietal2018}
{Stangalini}, M., {Jafarzadeh}, S., {Ermolli}, I., {et~al.} 2018, \apj, 869,
  110, \dodoi{10.3847/1538-4357/aaec7b}

\bibitem[{{Stangalini} {et~al.}(2021{\natexlab{a}}){Stangalini}, {Baker},
  {Valori}, {Jess}, {Jafarzadeh}, {Murabito}, {To}, {Brooks}, {Ermolli},
  {Giorgi}, \& {MacBride}}]{Stangalinietal2021a}
{Stangalini}, M., {Baker}, D., {Valori}, G., {et~al.} 2021{\natexlab{a}},
  Philosophical Transactions of the Royal Society of London Series A, 379,
  20200216, \dodoi{10.1098/rsta.2020.0216}

\bibitem[{{Stangalini} {et~al.}(2021{\natexlab{b}}){Stangalini}, {Jess},
  {Verth}, {Fedun}, {Fleck}, {Jafarzadeh}, {Keys}, {Murabito}, {Calchetti},
  {Aldhafeeri}, {Berrilli}, {Del Moro}, {Jefferies}, {Terradas}, \&
  {Soler}}]{Stangalinietal2021b}
{Stangalini}, M., {Jess}, D.~B., {Verth}, G., {et~al.} 2021{\natexlab{b}},
  \aap, 649, A169, \dodoi{10.1051/0004-6361/202140429}

\bibitem[{{Teriaca} {et~al.}(2008){Teriaca}, {Curdt}, \&
  {Solanki}}]{Teriacaetal2008}
{Teriaca}, L., {Curdt}, W., \& {Solanki}, S.~K. 2008, \aap, 491, L5,
  \dodoi{10.1051/0004-6361:200810209}

\bibitem[{{Thomas} {et~al.}(1984){Thomas}, {Cram}, \& {Nye}}]{Thomas1984}
{Thomas}, J.~H., {Cram}, L.~E., \& {Nye}, A.~H. 1984, \apj, 285, 368,
  \dodoi{10.1086/162514}

\bibitem[{{Thomas} \& {Weiss}(2012)}]{Thomas+Weiss2012}
{Thomas}, J.~H., \& {Weiss}, N.~O. 2012, {Sunspots and Starspots} ({Cambridge,
  UK}: {Cambridge Univ.\ Press}), \dodoi{10.1017/CBO9780511536342}

\bibitem[{{Weiss} \& {Proctor}(2014)}]{Weiss+Proctor2014}
{Weiss}, N.~O., \& {Proctor}, M.~R.~E. 2014, {Magnetoconvection} (Cambridge,
  UK: {Cambridge Univ.\ Press}), \dodoi{10.1017/CBO9780511667459}

\bibitem[{{Yurchyshyn} {et~al.}(2014){Yurchyshyn}, {Abramenko}, {Kosovichev},
  \& {Goode}}]{Yurchyshyn2014}
{Yurchyshyn}, V., {Abramenko}, V., {Kosovichev}, A., \& {Goode}, P. 2014, \apj,
  787, 58, \dodoi{10.1088/0004-637X/787/1/58}

\bibitem[{{Yurchyshyn} {et~al.}(2020){Yurchyshyn}, {Kilcik}, {{\c{S}}ahin},
  {Abramenko}, \& {Lim}}]{Yurch2020}
{Yurchyshyn}, V., {Kilcik}, A., {{\c{S}}ahin}, S., {Abramenko}, V., \& {Lim},
  E.-K. 2020, \apj, 896, 150, \dodoi{10.3847/1538-4357/ab91b8}

\end{thebibliography}

\end{document}